\RequirePackage{fix-cm}
\documentclass{svjour3}                     
\smartqed  
\usepackage{amsmath}
\usepackage{graphics}
\usepackage{soul}
\usepackage{graphicx}
\usepackage{amssymb}
\usepackage{braket}
\usepackage{dsfont}
\usepackage{slashed}
\usepackage[compat=1.0.0]{tikz-feynman}
\begin{document}

\title{Supersymmetry-Driven Quantum Gate Design Based on Feynman Path Integral and TPCP Map Optimization}

\author{Harish Parthasarathy{$^{1}$}, Monika Aggarwal{$^{2}$} and Kumar Gautam{$^{3^*}$}}


\institute{{$^{1}$} ECE division, Netaji Subhas University of Technology, India \\ {$^{2}$} Centre for Applied Research in Electronics (CARE), IIT Delhi, India\\ {$^{3}$} Quantum Computing Lab, QUANTUM RESEARCH AND CENTRE OF EXCELLENCE, New Delhi, India}

\maketitle

\begin{abstract}
We use supersymmetry to enlarge the dimension of the Hilbert space on which the unitary evolution of the state of the quantum fields acts. We discuss how to control the unitary evolution or TPCP maps generated by the quantum evolution of the fields by controlling the vacuum expectations of other fields in the theory. This amounts to breaking supersymmetry using control vacuum expectation values of the other fields. The evolution of the wave functional or TPCP maps obtained by tracing out over other fields is based on the Feynman path integral formula for the fields. By using the methods of quantum stochastic filtering, we estimate the evolving state of the fields from non-demolition noise measurements and then design a family of TPCP maps evolving in time whose outputs match the estimated evolving state. In this way, we are able to simulate the evolution of the state of the quantum noisy fields. Direct matching of the designed TPCP map to output the evolving system state is not possible since there is no way by which we can determine the exact evolving state, we can only estimate it using non-demolition measurements. The family of designed TPCP maps can be based on using a simulated master equation with unknown parameters incorporated into the Hamiltonian and the other Lindblad operators, chosen so as to match the state outputted by the quantum filter.

\end{abstract}

\section{Introduction}\label{s1}

The quantum mechanical symmetry known as supersymmetry (SUSY) connects the space-time continuum and the intrinsic degrees of freedom of fundamental particles. The Standard Model's hierarchy problem and ultraviolet divergence can be resolved in a SUSY theory where fermions and bosons have the same mass and cancel each other out in the Higgs particle's Lagrangian. Since the cancellation in Higgs mass is imprecise and no known particles have SUSY partners at present energy scales, SUSY must have spontaneously broken. It is very hard to figure out mathematically whether SUSY breaks on its own in quantum field theory. To help us understand this important property, supersymmetric quantum mechanics (SUSY QM) has been suggested as a play model. The boson-fermion correspondence is what makes all levels in a SUSY QM model degenerate, and Hamiltonians don't have negative eigenvalues. Bosonic or fermionic ground states may, on the other hand, exist. These supercharged vacuum states, which do not necessarily possess the boson-fermion mapping, serve as the origins of SUSY. Even though the SUSY theory has many applications in optics, condensed matter physics, quantum chaos, and even non-physical fields, cutting-edge high-energy physics experiments or other indirect experimental evidence have not demonstrated that it accurately describes the physical world. Still, many theoretical proposals exist for using quantum simulation to examine its effect on gate design. Ion traps are a top choice for quantum information processing due to their long coherence time, ease of setup and reading, and precise controllability with a microwave or laser. A number of quantum field theories, quantum phase transitions, and many-body dynamics simulations have been fruitful using these.

Based on the Feynman path integral, we propose a SUSY-driven quantum gate design and show that it spontaneously breaks SUSY in this model. It was theoretical physicist Richard Feynman who initially brought up the concept of quantum computing in 1982 \cite{Ref1}. Forty years of hard work have paid off in the areas of quantum computation, quantum simulation, and quantum information, but modern research in the area has exhibited new trends. In recent experiments, it has been demonstrated that increasing the number of qubits to dozens can result in achieving quantum dominance \cite{Ref1}-\cite{Ref4}. We can simulate quantum bits with this, since it is close enough to most classical computers. The fidelity of quantum logic gates and the quantum computing threshold value for fault tolerance both have a long way to go, but experimental platforms are steadily approaching the thousands of bits required for practical quantum computing. Due to their role as unitary operators for the evolution of quantum states within the context of Hilbert space, quantum gates serve as the fundamental building blocks of quantum algorithms. In the absence of universal quantum gates, quantum computation of any type would be mathematically impossible. Capacity to scale up to more qubits is of the utmost importance. Experts widely anticipate that the field of quantum computing and quantum information will experience unprecedented growth in the near future. Scientists and engineers must continue to explore the boundaries of this exciting new area if widespread implementation of quantum computing is to be achieved. There is a lot of hope that quantum computing can help with two areas: creating efficient algorithms to solve computational problems and accurately simulating physical systems. The underpinning of these computers is the quantum mechanical universe explanation. It is possible to solve the factoring problem exponentially faster and the searching problem quadratically faster using quantum mechanics \cite{Ref5}-\cite{Ref7}.. Additionally, highly effective quantum algorithms for solving linear systems of equations have been developed.


\par  The "qubit" is the fundamental unit of measurement and control in quantum computing. It is associated with two fundamental states of a quantum system and defines quantum superpositions. Any other transformation in the same domain can be applied to a universal transformation. In order to preserve the inner product of quantum states and their normalization conditions, only unitary quantum gates are required. A second physical specification of quantum mechanics for quantum state transformations is that the evolution from the initial state to the final state must be reversible.  Because the identity, $\mathds{1}$, can be obtained by applying $\mathbf{U^{\dagger}}$ to a unitary $\mathbf{U}$, unitary gates are reversible. Experiments using ion traps, nuclear magnetic resonance, and superconducting qubits have proven to be viable physical implementations of quantum gates. But large-scale quantum computers are still difficult to physically implement \cite{Ref8}- \cite{Ref10}..

\par The novelty of this paper is as follows: First, exploit additional fields provided by super partners to increase the Hilbert space dimensions on which the quantum gate acts. This approach allows for more complex quantum operations to be performed, potentially leading to improved computational capabilities. By leveraging the extra information from super partners, the quantum gate can achieve higher levels of precision and efficiency in its operations. Second, use symmetry breaking in physics as a natural method for providing control parameters to design the gate. Symmetry breaking in physics can introduce new degrees of freedom that can be harnessed to fine-tune the quantum gate's behavior. This innovative approach offers a promising avenue for enhancing the gate's performance and expanding its range of applications in quantum computing.  Third, design TPCP maps by averaging over non-observable fields to design TPCP maps arising due to the presence of quantum noise in an otherwise physical system described by unitary dynamics. Introduction of additional fields provided by supesymmetry enables us to account for noise. By incorporating these additional fields, we can effectively mitigate the impact of quantum noise on the system's performance, ultimately leading to more reliable and efficient TPCP maps. This approach not only improves the gate's overall functionality but also opens up new possibilities for utilizing quantum computing in various practical applications. And fourth, in a supersymmetric theory of nature, we can control the Gaugino vacuum expectation to make sure that the dynamics of the observable gauge field will follow a desired unitary family of gates. This is because gauge fields and gauge fields are naturally paired with each other. By manipulating the Gaugino vacuum expectation in this way, we can enhance the stability and accuracy of quantum gates in practical quantum computing scenarios. This innovative approach demonstrates the potential for harnessing supersymmetric theories to optimize quantum computing systems. 

\par This paper's main contribution is that it uses both operator theoretical Hamiltonian quantum mechanics and Feynman path interaction theoretical mechanics to create unitary gates and TPCP maps. This is done by controlling the expected value of upper partneers in the vacuum after averaging auxiliary fields. The auxiliary field is divided into two parts, one unobserbable super partner and the second control super partner, by averaging over unobserbable superpartnes. Unitary  maps evolved into TPCP maps that can be manipulated by controlling the vacuum expectation of controlled superpartnres. The combination of these two theoretical frameworks allows for a more comprehensive approach to creating unitary gates and TPCP maps. By incorporating both operator theoretical Hamiltonian quantum mechanics and Feynman path interaction theoretical mechanics, this paper presents a novel method for manipulating quantum systems with greater precision and control, shown in Fig 1. 

\par The main practical application of this gate is to design a very large dimension of Hilbert gate, which can then be used to design very large dimension QFT maps when noise is present in the system. This allows for the application of quantum mechanics to spectral analysis problems in classical signal processing and the establishment of a secure communication system. Furthermore, this method has the potential to significantly improve the efficiency of quantum algorithms and simulations, making it a valuable tool for researchers in various fields. Additionally, the integration of these two theoretical frameworks opens up new possibilities for exploring complex quantum phenomena and developing advanced quantum technologies. 

\par This paper presents the construction of a quantum gate and optimization of the TCPC map through the application of quantum electrodynamics, the quantum field theory of electron-photon interactions. If we assume that the electron is scattered from a random potential, we can find its potential by finding the minimum energy required to minimize the error between the actual gate and a specified unitary gate \cite{Ref11}\cite{Ref15}. When energy levels are excessively high, minimization fails. After discretization, the resulting equations are transformed to matrix equations and solved. It is a notable contribution to the field of quantum computing because the optimization process used in this design can be applied to other designs of quantum gates \cite{Ref21}. The fact that the path optimization is mostly unknown adds a great deal of complexity to the problem. Careful decision-making is required to advance technology, which is the key to solving this problem. You can use deterministic, heuristic, and metaheuristic algorithms to solve the route optimization problem. Each technique has its own set of requirements and restrictions, but they can all solve a small number of straightforward path problems \cite{Ref27}.

\par The rest of the paper is structured as follows: in section 2, we lay the groundwork for quantum electrodynamics and talk about how to formulate problems. Section 3 details the quantum gate model that has been proposed, as well as the design of the unitary gate and the TPCP map, which are based on the Feynman path integral. The outcome of the simulation, which needs to be limited, is derived in Section 4. The conclusion and discussion of what comes next are found in Sections 5.
 
\section{Problem formulation and discussion}\label{s3}

By assuming that the laws of nature are inherently supersymmetric in nature, ie, invariant under Boson-Fermion exchange, we are able to introduce additional quantum fields into the action leading to more degrees of freedom in the corresponding Hamiltonian, ie, an enlargement of the dimension of the underlying Hilbert space obtained by truncation of the Boson and Fermion Fock spaces \cite{Ref16}-\cite{Ref19}. This amounts to setting up a supersymmetric action and hence a supersymmetric Hamiltonian using Chiral superfields for describing matter, non-Abelian gauge superfields for describing the particles exchanged in the interaction of the matter superfields. The total Lagrangian density thus has the form $[\Phi^*.exp(t.V).\Phi]_D+Re[f(\Phi)]_F+Re[W_L^T\epsilon.W_L]_F$. Here, $\Phi$ is the left Chiral matter superfield whose components are the scalar field, the left Chiral Dirac field and an auxiliary $F$-field. $V$ is the non-Abelian gauge superfield whose components in the Wess-Zumino gauge are the Yang-Mills gauge field, the gaugino field and an auxiliary $D$ field. The first term in the above Lagrangian is the supersymmetric generalization of the standard Lagrangian of the Klein-Gordon scalar field plus the Lagrangian of the massless Dirac field plus the interaction between the Dirac field and the non-Abelian Yang-Mills gauge field. The second term is a superpotential term that after eliminating the auxiliary F and D fields using  Gaussian path integration, contains a term that corresponds to a mass term of the Dirac field whose mass depends on vacuum expectation value of the scalar field via the superpotential and the last term is the supersymmetric generalization of the Lagrangian density of the Yang-Mills non-Abelian gauge field.
Now the crucial step here is to introduce control parameters $\beta$ into the superpotential $f(\Phi)$ and hence eventhough these parameters can be varied, supersymmetry of the total Lagrangian will not be broken and hence the laws of nature will remain preserved. By varying these parameters, we can therefore hope to control the unitary evolution of the wave functional of the fields $\phi,\psi,V^A_{\mu}, \lambda^A$ starting from $t=-\infty$ and going $upto t=+\infty$ in the scattering process. Here, $\phi,\psi,V^A_{\mu},\lambda_A$ are respectively the scalar field, the Dirac field, the non-Abelian gauge field and the gaugino field. We call $\psi$ as the Higgsino field, ie, theFermionic superpartner of the Bosonic Higgs field $\phi$ because under an infinitesimal supersymmetry transformation defined by the Salam-Strathdhee supervector field
$$
\alpha^TL, L=\gamma^{\mu}\theta.\partial_{\mu}+\gamma^5\epsilon.\partial_{\theta}
$$
applied to the left Chiral superfield $\Phi$, it becomes clear that the variation of $\phi$ is proportional to $\psi$ and the variation of $\psi$ is proportional to $\phi$ plus the auxiliary $F$ field. Likewise, the gaugino field $\lambda^A$ is the Fermionic superpartner of the Bosonic gauge field $V^A_{\mu}$ since these are exchanged under the action of the above supervector field. Variation of the parameters of the superpotential amounts after eliminating the auxiliary fields, to controlling the mass of the Dirac field via the scalar field. There are two ways to design the quantum gate as either a unitary map or a TPCP map, one based on Feynman's path integral method and the other based on Hamiltonian quantum mechanics using creation and annihilation operators. Here, we discuss both of these methods.

After eliminating the auxiliary $F$ and $D$ fields, the total Lagrangian density has the form
$$
\mathcal L(\phi,\psi,\lambda,V)=(D_{\mu}\phi)^*(D^{\mu}\phi)+\psi^*\gamma^0(\gamma^{\mu}.(i\partial_{\mu}+gV^A_{\mu}t_A)-G(\phi))\psi-(1/4)F^{A\mu\nu}F^A_{\mu\nu}
$$
$$
+k(1).\lambda^*\gamma^0(\gamma^{\mu}(i\partial_{\mu}+gV^A_{\mu}ad(t_A))\lambda
$$
$$
+F_1(\psi,\lambda,\phi)+F_2(\phi)
$$,
where in the second term, $G(\phi)$, the Dirac mass matrix is proportional to $f''(\phi)$ and where the second last term $F_1$ is trilinear in $(\psi,\psi^*),(\lambda,\lambda^*),(\phi,\phi^*)$. The last term $F_2$ is proportional to $|f'(\phi)|^2$. It comes mainly from the observation that $[\Phi^*\Phi]_4$ contains a term $F^*F$, while $Re[f(\Phi)]_2$ contains a term $Re(f'(\phi)F)$, so by eliminating $F$, setting the variation of the action w.r.t it to zero, we get $F=-f'(\phi)^*/2$ and substituting this back into the same expressions, we get a contribution
$$
F^*F+Re(f'(\phi)F)=-|f'(\phi)|^2/4
$$
It easily follows that the effective potential of the supersymmetric action is $|f'(\phi)|^2/2$ so that for the vacuum to be supersymmetric, we require
that $f'(\phi_0)=0$ where $\phi_0$ is the vacuum expected value of the scalar field. Therefore, if we do not wish to break supersymmetry but simultaneously require control of the superpotential via the parameters $\beta$, we must then move on the p-dimensional surface $f'(\phi_0|\beta_1,...,\beta_p)=0$ in the parameter space. Usually, the superpotential is gauge invariant, ie,
$f'(\phi).t_A\phi=0$ forall $\phi$ where $t_A$ runs over the gauge group generators. If $d$ is the dimension of the gauge group, this equation imposes $d$ constraints on the super-potential. Therefore, the effective number of constraints in the equation $f'(\phi_0|\beta)=0$ is $n-d$ where $n$ is the number of $\phi$-components. This means that the allowable dimension of the parameter manifold for varying $\beta$ is $p-(n-d)$ for designing the optimally controlled gate \cite{Ref20}-\cite{Ref25}.
\bigskip

In this expression,
$$
D_{\mu}\phi=(\partial_{\mu}-igV^A_{\mu}t_A)\phi
$$
Note that $\psi$ transforms according to the vector representation of the gauge group while $\lambda=\lambda^At_A$ transforms according to the adjoint representation of the gauge group. Now from this Lagrangian, we can write down the equations of motion and obtain the free wave (unperturbed) solution components for the fields $\phi,\psi,\lambda^A,V_{\mu}^A$ in terms of Bosonic and Fermionic creation and annihilation operators. Specifically, the unperturbed equations are
$$
\partial^{\mu}\partial_{\mu}\phi=0,\gamma^{\mu}\partial_{\mu}\psi=0, \gamma^{\mu}\partial_{\mu}\lambda^A=0, 
$$
$$
(\delta^{\mu}_{\rho}\partial^{\nu}\partial_{\nu}-\partial^{\mu}\partial_{\rho})V^A_{\rho}=0,
$$
The solutions to these equations are plane waves with coefficients being Bosonic and Fermionic creation and annihilation operators in momentum space.
We denote these unperturbed solutions by $\phi^0,\psi^0,\lambda^{0A}, V^{A0}_{\mu}$. These solutions represent the evolving observables in Dirac's interaction picture wherein observables evolve according to the unperturbed Hamiltonian while states evolve according to the perturbation in the Hamiltonian followed by a rotation using the adjoint representation of the unperturbed Hamiltonian \cite{Ref36}-\cite{Ref40}.
\bigskip

\section{Unitary gate and TPCP map design based on the Feynman path integral}

We write down the total action functional for the Chiral and gauge superfields interacting with each other as
$$
S[\phi,\psi,V,\lambda]=S_{01}(\phi|\beta)+S_{02}(\psi)+S_{03}(V)+S_{04}(\lambda)
$$
$$
+S_{11}(\phi,V)+S_{12}(\psi,V)+S_{14}(\lambda,V)+S_{15}(\psi,\phi|\beta)+S_{16}(\phi,\psi,\lambda|\beta)
$$
with obvious meanings for the various terms. The superpotential terms are present in $S_{01}$ and also in $S_{15}$ and $S_{16}$. These terms arise when we eliminate the auxiliary $F$ and $D$ by setting the variational derivative of the action w.r.t these to zero. This is justified because the auxiliary fields enter into the Lagrangian linearly and quadratically, so the effective action obtained by path integrating the exponential of the total action w.r.t these fields is a Gaussian functional integral and hence can be evaluated by setting these fields to their values at which the action is stationary w.r.t them.

Now, suppose that we have detected the superpartners $\lambda$ and $\phi$ of the standard fields $V_{\mu},\psi$, namely the superpartners of the fields used in standard quantum electrodymanics. Then, we should using path integrals be able to write down the unitary evolution kernel for the three fields simply by path integrating $exp(iS)$ over all the fields from time $t=-\infty$ to $t=+\infty$ with specified spatial values of these fields at times $t=-\pm\infty$. On the other hand, if as is the present case, we have not detected the superpartners $\lambda,\phi$ of the standard qed fields $V_{\mu},\psi$. Then, by path integrating the evolution kernel in the adjoint domain over these superpartners, we would get a non-unitary TPCP map that transforms an initial pure or mixed state of the fields $(\psi,V^A_{\mu})$ to a mixed state of the same fields. In both cases, we can control the $\beta$ parameters to design either a unitary gate or a TPCP gate that is as close as possible w.r.t some distance measure to a given unitary or TPCP gate. Of course, the TPCP gate acts in a lower dimensional Fock space as compared to the original unitary gate because it is obtained by path integrating, which amounts to partial tracing over the scalar and gaugino fields \cite{Ref26}-\cite{Ref29}.

It should be noted that when the vacuum has finite positive energy, then by applying a supersymmetry transformation to it using one of the supersymmetry generators $Q$, we can transform a Bosonic state to a Fermionic state of the same energy and vice versa. However, the condition for broken supersymmetry is that the vacuum has nonzero positive energy. Thus, broken supersymmetry means that the vacuum state has positive energy and hence that a Bosonic vacuum state can be paired with a Fermionic vacuum state and vice versa. Therefore, unbroken supersymmetry which implies that the vacuum energy is zero also means that the vacuum cannot be paired with any other state and is therefore invariant under the supersymmetry generators $Q$. A nice discussion of this account of unbroken and Broken supersymmetry can be found in [Steven Weinberg, vol.III, Supersymmetry] where it is also mentioned that if $F$ is the operator that has eigenvalue zero on a Bosonic state and eigenvalue one on a Fermionic states, so that $(-1)^F$ has eigenvalue $1$ on Bosonic states and eigenvalue $-1$ on Fermionic states. Then, the Witten index $Tr((-1)^F)$ will get contributions only from zero energy states, because positive energy states of Bosons and Fermions pair up leading to an equal number of Bosonic and Fermionic states at any fixed positive energy \cite{Ref30}. It follows that $Tr((-1)^F)$ is non-zero and equal to the number of Bosons minus the number of Fermions in the vacuum. In particular, supersymmetry is unbroken iff the vacuum has zero energy, iff Bosonic and Fermionic states in the vacuum cannot be perfectly paired. This amounts to saying that if supersymmetry is unbroken, then we can have a vacuum state that is either purely Bosonic without any Fermion to pair up with these Bosons or vice versa.

\section{Simulation Results}

Super Yang-Mills theory is the simplest supersymmetric field model having just one gauge Boson and one gaugino field which are super-partners of each other. This can be derived from the discussion at the beginning of this article by deleting the scalar Higgs field $\phi$ and its super-partner, the Dirac Higgsino field $\psi$. Thus, the there are only two component fields,
namely the gauge field $V^A_{\mu}$ and the gaugino field $\lambda^A$. This theory can be derived from the super-symmetric Lagrangian \cite{Ref31}-\cite{Ref35}.
$[W_A^T\epsilon W_A]_2$ where $W_A$ is the left Chiral spinor superfield defined by
$$
W_A=D_R^T\epsilon D_Rexp(-t.V)(D_Lexp(t.V))
$$
with
$$
V(x,\theta)=V^A(x,\theta)t_A
$$
$V^A$ being the gauge superfield. This Lagrangian contains a $D^2$ term which can be eliminated by setting it by noting that path integration sets it to zero, its value at which the Lagrangian is stationary. More generally, we can add to this Lagrangian a supersymmetric term $\xi_AD^A$ linear in the $D$ and then path integration sets $D^A$ to a constant. 

Here, we simulate a quantum unitary gate and a TPCP map based on the super Yang-Mills theory. The Lagrangian in this theory is
$$
L=(-1/4)F^{A\mu\nu}F^A_{\mu\nu}+(\lambda^A)^T\gamma^5\epsilon\gamma^{\mu}[D_{\mu},\lambda^A]
$$
where
$$
D_{\mu}=\partial_{\mu}+ig.V^A_{\mu}t_A
$$
$$
[t_A,t_B]=-iC(ABC)t_C
$$
and
$$
igF^A_{\mu\nu}=[D_{\mu},D_{\nu}]^A
$$
or equivalently,
$$
F^A_{\mu\nu}=V^A_{\nu,\mu}-V^A_{\mu,\nu}+g.C(ABC)V^B_{\mu}V^C_{\nu}
$$
It is easily seen that by adopting the gauge condition $V^A_0=0$, with the canonical position fields as $Q^A_rV^A_r, r=1,2,3$ so that the canonical momentum fields for the gauge part of the action become
$$
P^A_r=\partial L/\partial V^A_{r,0}=F^A_{0r}
$$
we can express the corresponding Hamiltonian by applying the Legendre transformation in the form $(1/2)F^A_{0r}F^A_{0r}+(1/4)F^A_{rs}F^A_{rs}$, or equivalently in abbreviated notation as
$$
H_{gge}=C_1(rs)P_rP_s+C_2(rs)Q_rQ_s+C_3(rsk)Q_rQ_sQ_k+C_4(rskm)Q_rQ_sQ_kQ_m
$$
Here, $Q_r,P_r$ are abbreviated notations for the canonical position and momentum variables of the gauge part of the action. The Hamiltonian of the gaugino part of the action, on the other hand is the sum of a free gaugino part $i(\lambda^A)^T\gamma^5\epsilon\gamma^r\partial_r\lambda^A$ and an interaction part with the gauge potential $C(ABC)(\lambda^A)^T\gamma^5\epsilon\gamma^rV^B_r\lambda^C$. Denoting by $q_r,p_r$ the canonical position and momentum variables of this gaugino part of the action, we can express the gaugino Hamiltonian along with its interaction with the gauge part as
$$
H_{gino}=D_1(rs)p_rq_s+D_2(rsk)p_rq_kQ_s
$$
Note that $Q_r's$ are built from $V^A_r$, $P_r's$ from $V^A_{r,0}$, $q_r$ from $\lambda^A$ and finally, $p_r$ from $(\lambda^A)^*\gamma^0=(\lambda^A)^T\gamma^5\epsilon$ using the fact that the $\lambda^A$ are Majorana Fermions. Now comes the crucial supersymmetry breaking argument: When the gaugino fields acquire vacuum expectation values, the second term in $H_{gino}$ gets replaced by $D_2(rsk)<p_r><q_k>Q_s$ which can be expressed as $a(s)Q_s$ where
$a(s)=D_2(rsk)<p_r><q_k>$ are parameters dependent upon the vacuum expectations of the gaugino field and this amounts to the following effective Hamiltonian for the gauge field:
$$
H_{gge,eff}(Q,P)=C_1(rs)P_rP_s+C_2(rs)Q_rQ_s+C_3(rsk)Q_rQ_sQ_k+C_4(rskm)Q_rQ_sQ_kQ_m+a(s)Q_s
$$
We shall base our simulation studies on this model.

To simplify matters further, we shall assume that there is just one position variable for the gauge field and therefore the controlled Hamiltonian can be expressed as
$$
H(Q,P|a(t))=(P^2+Q^2)/2+c(1)Q^3+c(2)Q^4+b(t)Q
$$
where we have allowed the parameter $b=b(t)$ to depend on time. This looks like the Hamiltonian of a one dimensional Harmonic oscillator with cubic and fourth degree anharmonic terms plus a control electric field interaction term assuming that the harmonic oscillator particle carries charge. Thus, the control parameter $b(t)$ can be interpreted as a time varying control electric field. Introduce Using time independent perturbation theory and the quantum theory of the harmonic oscillator, we can compute the approximate eigenfunctions $|u_n>, n\geq 0$ and corresponding energy eigenvalues $E_n, n\geq 0$ of the Harmonic oscillator with anharmonic perturbations described by the Hamiltonian
$$
H_0=(P^2+Q^2)/2+c(1)Q^3+c(2)Q^4
$$
Then, using time dependent perturbation theory, we can compute the approximate evolution operator of the perturbed system as
$$
U(T)\approx U_0(T)-i\int_0^Tb(t)U_0(T-t)QU_0(t)dt
$$
where
$$
U_0(t)=exp(-itH_0)=\sum_nexp(-itE_n)|u_n><u_n|
$$
Truncation to $N+1$ dimensions gives the $N+1\times N+1$ dimensional unitary gate
$$
U_N(T)=U_{0N}(T)-i\sum_{n,m=0}^N\int_0^Tb(t)exp(-iE_n(T-t)-iE_mt)<u_n|QW|u_m>|u_n><u_m|
$$
$$
=\sum_{n,m=0}^N[\delta[n-m]-\int_0^Tb(t)exp(i(E_n-E_m)t)dt)exp(-iE_nT)<u_n|Q|u_m>]|u_n><u_m|
$$
and we can control the function $b(t)$ or better still, its truncated Fourier transform $\hat b_T(\omega)=\int_0^Tb(t)exp(i\omega t)dt$ evaluated at $E_n-E_m$ so that the gate $U_N(T)$ having matrix elements
$$
<u_n|U_N(T)|u_m>=
$$
$$
\delta[n-m]-\int_0^Tb(t)exp(i(E_n-E_m)t)dt)exp(-iE_nT)<u_n|Q|u_m>
$$
($n,m=0,1,..., N$) is as close as possible to a desired $N+1\times N+1$ unitary gate w.r.t the Frobenius norm. The goal is to minimize the difference between the gate $U_N(T)$ and the desired unitary gate in terms of the Frobenius norm, ensuring accuracy in quantum computing operations. By controlling the function $b(t)$ or its truncated Fourier transform, we can optimize the performance of $U_N(T)$ for various quantum computational tasks. This optimization process allows for more precise and efficient quantum computations, ultimately enhancing the capabilities of quantum computing systems. The ability to fine-tune $U_N(T)$ through adjustments in $b(t)$ provides a versatile approach to achieving desired outcomes in quantum algorithms and simulations. 
\bigskip

\section{Conclusion Discussion of Future Work}

We will discuss the Fermionic filter in the formalism in subsequent work. The Fermionic filter plays a crucial role in particle physics experiments, helping to distinguish between different types of particles based on their spin. Understanding its formalism is essential for accurately interpreting experimental results and making new discoveries in the field. 

Let $A(t)$ be the Bosonic annihilation process and $J(t)$ the Fermionic annihilation process.
Let $\Lambda_A(t)$ be the associated Bosonic counting process and $\Lambda_J(t)$ the associated Fermionic counting process. The system Hilbert space $\mathcal h$ is assumed to be
$\Bbb Z_2$ graded:
$$
\mathfrak h=\mathfrak h_0\oplus\mathfrak h_1
$$
Let $P_0$ denote the projection onto $\mathcal h_0$ and $\mathfrak h_1$ the projection onto $\mathfrak h_1$. It is clear that for describing unitary evolution, Bosonic noise must be coupled to the even system operators while Fermionic noise must be coupled to odd system operators. Note that $X$ is an even system operator iff
$$
X(\mathfrak h_0)\subset\mathfrak h_0, X(\mathfrak h_1)\subset\mathfrak h_1,
$$
and it is an odd system operator iff
$$
X(\mathfrak h_0)\subset\mathfrak h_1, X(\mathfrak h_1)\subset\mathfrak h_0
$$
Also note that $P_0+P_1=I$ and that if $X$ is any system operator, then
$$
X=X_0+X_1
$$
where $X_0$ and $X_1$ are respectively even and odd system operators. They are given by
$$
X_0=P_0XP_0+P_1XP_1, X_1=P_0XP_1+P_1XP_0
$$
Define the linear map $\tau$ on the space of system operators by the equation
$$
\tau(X)=X_0-X_1=(P_0-P_1)X(P_0-P_1)=\theta.X.\theta
$$
where
$$
\theta=P_0-P_1
$$

\section*{Author contributions}
The theoretical interpretation, writing, and review of the manuscript were done by H.P., M.A., who also revised the paper and included the quantum gate design, and K.G., who elaborated on some concepts and validated them through simulation.

\section*{ACKNOWLEDGEMENT}
Prof. KR Parthasathy sir provided invaluable guidance and support during this work, and we are grateful to IIT Delhi for providing the necessary facilities. In order to accomplish our research objectives, his guidance and knowledge were crucial. 

\section*{Funding}
No Funding 

 \section*{Data availability statement }
 No data availability 

\section*{Conflicts of interests}
On behalf of author, the corresponding author states that there is no conflict of interest.

\section*{Additional information}
Correspondence and requests for any materials should be addressed to K.G.

%
%
{}

\end{document}